\input amstex
\documentstyle{amsppt}

\centerline {SUBMODELS OF THE SPECIAL COMPRESSIBLE FLUID ON TWO-DIMENSIONAL
SUBALGEBRAS}

\centerline { 1. Introduction}

The differential equation of gas dynamic (EGD) are:
$$D=\partial_t+{ u}\cdot\nabla$$
$$D{ u}+\frac{1}{\rho}{\nabla}p=0,$$
$$D\rho+{\rho}{ div}{{ u}}=0, \eqno (1)$$
$$DS=0,$$
where ${ u}$ is speed, $\rho$ is density, $p$ is pressure, $S$
is entrophy. These variables with the special state equation admit
11 parametrical algebra operators Lee $L_{11}$. In the Cartesian
system of coordinates the $L_{11}$ bases is the following [1, see
also 2]:

$ X_1=\partial_x,\ X_2=\partial_y,\
X_3=\partial_z,X_4=t\partial_x+\partial_u,\
X_5=t\partial_y+\partial_v$,

$ X_6=t\partial_z+\partial_w,\ $ $
X_7=y\partial_z-z\partial_y+v\partial_w-w\partial_v, $

$ X_8=z\partial_x-x\partial_z+w\partial_u-u\partial_w, $ $
X_9=x\partial_y-y\partial_x+u\partial_v-v\partial_u, $ $
X_{10}=\partial_t, $

$ X_{11}=t\partial_t+x\partial_x+y\partial_y+z\partial_z, $

We considered the special state equation of the kind:
$$p=\pm{\rho}^{\gamma}+F(S),\eqno (2)$$
where $+{\rho}^{\gamma}$ for $\gamma>0$ and $-{\rho}^{\gamma}$ for
$\gamma<0$, $\gamma$~---~параметр, $F(S)$ is the function of
entropy. It coordinates with the fixed state equation for the
fluid at high pressures and high temperatures.

EGD with the equation (2) admit additional operators:

- stretching $
X_{12}=t\partial_t-u\partial_u-v\partial_v-w\partial_w-
(\overline{\gamma}-2)\rho\partial_\rho-\overline{\gamma}
p\partial_p, $

- carry $ X_{13}=\partial_p, $
\newline where` $\overline{\gamma}=2\gamma/(\gamma-1)$, $\gamma\neq1$.
\newline Together with $L_{11}$ L they make up algebra Lee $L_{13}$.

In cylindrical coordinates (C) $ x=(x,r,\theta),
u=(U,V,W),  y=r\cos\theta,  z=r\sin\theta, u=U,
v=V\cos\theta-W\sin\theta, w=V\sin\theta+W\cos\theta $ the basis
of algebra $L_{13}$ is the following: $ X_1=\partial_x $, $
X_2=\cos\theta\partial_r-{\sin\theta}{r^{-1}}(\partial_\theta+W\partial_V-
V\partial_W) $,
\newline $
X_3=\sin\theta\partial_r+{\cos\theta}{r^{-1}}(\partial_\theta+W\partial_V-
V\partial_W) $, $ X_4=t\partial_x+\partial_U $,
\newline $
X_5=\cos\theta(t\partial_r-\partial_V)-{\sin\theta}{r^{-1}}t
(\partial_\theta+W\partial_V-(V-{r}{t^{-1}})\partial_W) $,
\newline $
X_6=\sin\theta(t\partial_r+\partial_V)+{\cos\theta}{r^{-1}}t
(\partial_\theta+W\partial_V-(V-{r}{t^{-1}})\partial_W) $, $
X_7=\partial_\theta $,
\newline $
X_8=\sin\theta(r\partial_x-x\partial_r+V\partial_U-U\partial_V)+
\cos\theta(W\partial_U-U\partial_W-{x}{r^{-1}}
(\partial_\theta+W\partial_r-V\partial_W)) $,
\newline $
X_9=-\cos\theta(r\partial_x-x\partial_r+V\partial_U-U\partial_V)+
\sin\theta(W\partial_U-U\partial_W-{x}{r^{-1}}
(\partial_\theta+W\partial_V-V\partial_W))$ ,\newline
$X_{10}=\partial_t, X_{11}=t\partial_t+x\partial_x+r\partial_r $,
\newline $
X_{12}=t\partial_t-U\partial_U-V\partial_V-W\partial_W-
(\overline\gamma-2)\rho\partial_\rho-{\overline\gamma}p\partial_p,
X_{13}=\partial_p $.

For algebra $L_{13}$  all subalgebrus are listed  [4], If
\newline $\gamma=-1, 1/3$  there are more subalgebrus than for any
$\gamma$.
We shall considers two-dimentional subalgebrus from optimum system
for  $L_{13}$, appering only when $\gamma=-1, 1/3$. We shall
write out not similar two-dimentional subalgebrus for this
purprose:

$ 2.1'.    X_1+X_2, aX_4+X_{13}, a({\overline\gamma}-1)=0; $

$ 2.2'.   X_{12}, aX_4+X_{13}, a\neq0; $

$ 2.3'.   X_1+X_{12}, aX_4+bX_5+X_{13}; $

$ 2.4'.   -X_{11}+X_{12}, X_1+aX_5+X_{13}, a\neq0; $

$ 2.1''.  aX_1+X_{12}, X_{10}+X_{13}, a\neq0; $

$ 2.5'.   aX_7-bX_{11}+X_{12}, X_1+X_{13}, a\neq0; $

$ 2.6'.
aX_7+bX_{11}+X_{12}, cX_4+X_{13}, c^2+(b+1)^2\neq0{\vee}a^2+
c^2\neq0, c({\overline\gamma}-1)=0; $

$ 2.7'.   X_1+aX_7+X{12}, bX_4+X{13}, a\neq0;  \
(3)$

$ 2.8'.   aX_7+X_{12}, bX_4+X_{13}, a\neq0,
b(\overline\gamma-1)=0; $

$ 2.9'.
aX_7-(\overline\gamma+1)X_{11}+X_{12}, bX_4+X_{10}+X_{13},
b(\overline\gamma-1)=0, \overline\gamma\neq1; $

$ 2.2''.  bX_1+aX_7+X_{12}, X_{10}+X_{13}, b\neq0; $

$ 2.10'.  aX_7+X_{10}-X_{11}+X_{12}, bX_1+X_{13}, b\neq0; $
\newline are for subalgebrus  $2.1''$ and $2.2''$,
$\overline\gamma=-1\Rightarrow\gamma=1/3$,
\newline for other subalgebrals $\overline\gamma=1\Rightarrow\gamma=-1$,
\newline here parameters $a$ and $b$ which define set of series of
dissimilar subalgebras.

\centerline { 2. Proposal for the coordinates of the equation (2)
with the fixed state equation}

The equation (2) coordinates with the fixed state equation [5]
$$p=\Phi({\rho}^{-1})+Tf({\rho}^{-1}), \eqno (4)$$
for same implication functions $F(S)$, $\Phi(\rho^{-1})$,
$f(\rho^{-1})$. Where $\Phi({\rho}^{-1})$ is describes the
potential component of pressure, $Tf(\rho^{-1})$ is the describes
the thermal components of pressure, $\rho^{-1}$ -- is the specific
volume. The equation (4) describes behavior of the real
environments, which by their properties approximate firm or fluid
frames. It is possible with high pressure (about
$10^{9}$kg/{cm}$^2$) and high temperature (about  $10^6K$). Let us
find the meanings $F(S)$, $\Phi(\rho^{-1})$, $f(\rho^{-1})$.

Comparing $p$ в (2), (4) and excluding $T$ with the help of
thermodynamics ($\rho, S$ -- are independent parameters) we
receive the identity:
$$\pm\rho^{\gamma}+F(S)=\Phi(\rho^{-1})+(G'_S-F'_S{\rho}^{-1})f(\rho^{-1}),\
eqno
(5)$$ where $G(S)$ is determined by the additional experiment.
\newline After twice differenting by $S$:
$$0=-F'_S-F''_{SS}Vf(V)+G''_{SS}f(V), \eqno (6)$$
where $V=\rho^{-1}$.

${1^0.}$ Let us $F_{SS}\neq0$, then:
$$\frac{F'_S}{F''_{SS}}=-Vf(V)+\frac{G''_{SS}}{F''_{SS}}f(V).\eqno
(7)$$ Once again we differentiate on $S$ and receive:
$(\frac{F'_S}{F''_{SS}})'=(\frac{G''_{SS}}{F''_{SS}})'f(V)$. If
$\frac{G''_{SS}}{F''_{SS}}\neq0$, then, dividing by variables we
have $f$=const=$f_0$ and after the integration we subtitle it in
(7). We take the contradiction , that $\rho, S$ are independing
parameters.

Means $\frac{G''_{SS}}{F''_{SS}}=0$, i.e.
$$G''_{SS}=k_0F''_{SS}, F'_S=k_1F''{SS},\eqno (8)$$
and from (7) follows, that $k_1=-Vf(V)+k_0F(V)$.

The integration of (8) at $k_1\neq0$ and the substitution to (5)
gives:
$$
F(S)=k_1k_2e^{\frac{S}{k_1}}+k_3,
$$
$$
\Phi(\rho^{-1})=\pm\rho^\gamma+k_3-\frac{k_4k_1}{k_0-\rho^{-1}},\eqno
(9)$$
$$ f(\rho^{-1})=\frac{\rho{k_1}}{\rho{k_0}-1},
$$
$$
G(S)=k_0k_1k_2e^{\frac{S}{k_1}}+k_4S+k_5,
$$
where $k_j$ are the constants of integration.

$2^0$. Let us  $F_{SS}=0$ (it is equivalent $k_1=0$). Then
$F(S)=k_1S+k_0$ and from (6) we receive (if $G''_{SS}\neq0$)
$\frac{k_1}{G''_{SS}}=f(V)$=const=$f_0$. Hence from (5) are
follows:
$$
\Phi(\rho^{-1})=k_1\rho^{-1}f_0+k_0-k_2f_0\pm\rho^\gamma,
$$
$$f(\rho^{-1})=f_0,\eqno (10)$$
$$
G(S)=\frac{k_1}{2f_0}S^2+G_1(S)+G_0,
$$
where $G_0,G_1$=const.

$3^0$. Let us  $F_{SS}=0$, $G_{SS}=0$. Then from (7) follows,
that $F_S=0$, $F(S)=F_0$. And from (5) we shall receive:
$$
\Phi(\rho^{-1})=\pm\rho^\gamma+F_0-G_1f(\rho^{-1}),
$$
$$G(S)=G_1(S)+G_0,\eqno (11)$$
where $F_0$, $G_0$, $G_1$ are constants.

Thus, the state equation  (2) coordinates with the equation (4),
if the functions $F(S)$, $f(\rho^{-1})$, $\Phi(\rho^{-1})$ are
represents in one of kinds: (9), (10), (11).

\centerline { 3. Calculation of invariants}

For construction the submodels of the special compressible fluid
necessary to calculate invariants of the subalgebrus. [2, see also
1].

The algorithm of the calculation invariants consist in the
following:

1. We select the system of coordinates, in witch calculates the
invariants. If the subalgebra contains the operator $X_7$ of the
rotation, it is convenient to choose cylindrical coordinates. If
the operator of the rotation id not resent are convenient the
cartesian coordinates.

2. We write out the operators  of subalgebra in convenient system
of coordinates from the list (3).

3. We enter the function  $h$,  which depends from 9 variables
$(t,{ x},{ u}, {\rho},{p})$ as required invariants.

4. The function $h$ is invariants of the subalgebra $L=<Y_1,Y_2>$
Only when any operators $Y$ of subalgebra,
 working on invariants function, to annul it. Namely, $Y\cdot{h}=0,
Y\in{L}$. We shall work by the operator $Y_1$ of the basis
subalgebra $L$ on invariant function. In result we received the
linear homogeneous equation with the partial derivatives of the
1-st order. For this equation we write the characteristic
equation, the system of the ordinary differential equation [6].
Let us assume, that there is an obviously complete set
functionally independent invariant (integrals) $I^k(t, x,
u,\rho,p)$, $k=1..8$.

4. We write down the second operator of the basis through received
invariants by the rule:
$$Y_2=\xi^j\partial_{x^j}=\xi_j\frac{\partial{I^k}}{\partial{x^j}}\partial_{
I^k}\eqno
(12)$$

5. We shall work by the stayed operator $Y_2$ on invariant
function $h(I^k)$. We received the linear homogeneous equation
with the partial derivatives of the 1-st order. We write down for
it the equation of the characteristic. We find a complete set of
invariants.

6. We pass to initial variable.

The received invariants (3) are shown into the table (see the
appendix).

{ Example:}

As an example we shall consider the subalgebra  $2.7'$ from (3):
\newline $Y_1=X_1+aX_7+X_{12}=a\partial_{\theta}+t\partial_t-U\partial_U-
V\partial_V-W\partial_W+\rho\partial_{\rho}-p\partial_p$,
\newline $Y_2=bX_4+X_{13}=bt\partial_x+b\partial_U+\partial_p$.
\newline Let us the invariant function $h(t,{ x},{ u},{\rho},{p})$,
${ x}=(x,r,\theta)$, ${ u}=(U,V,W)$,
\newline satisfying to the equations  $Y_1{\cdot}h=0, Y_2{\cdot}h=0$.
\newline The second equation looks like $bth_x+bh_U+h_p=0$.
\newline ЗLet us write down the equation of the characteristics:
$${\frac{dx}{bt}}={\frac{dU}{b}}={\frac{dp}{1}}=
{\frac{dr}{0}}={\frac{d{\theta}}{0}}=
{\frac{dV}{0}}={\frac{dW}{0}}=
{\frac{d{\rho}}{0}}={\frac{dt}{0}}.$$ We find integrals, which
form a complete set functionally independent invariants $\colon$
$t;{\rho};W;V;{\theta};r;U_1=U-xt^{-1};$ $p_1=p-x(bt)^{-1}.$

We shall receive $h_{1x}=0$, when written down the second equation
through invariants by a rule (12). Means
$h=h_1(t,r,{\theta},V,W,{\rho},p_1,U_1).$

The first equation looks like in new invariants for the  known
equations variable $\colon$
$$ah_{1{\theta}}\!\!+\!\!{t}h_{1t}\!\!+\!\!(-U\!\!+\!\!xt^{-1})h_{1U_1}\!\!-
\!\!Vh_{1V}\!\!-\!\!Wh_{1W}\!\!+\!\!{\rho}h_{1{\rho}}\!\!+
\!\!(-p\!\!+\!\!x(bt)^{-1})h_{1p_1}\!\!=\!\!0;$$
\par Having written down the characteristic equation
and having calculated integrals, we receive a complete set
functionally independent invariants, which in initial variable
look like $\colon$
$$r;{\theta}-a{\ln}{\mid}t{\mid};Ut-x;Vt;Wt;{\rho}t^{-1};pt-xb^{-1}.
\eqno (13)$$

\centerline { 4. Invariant submodels of the second rank}

\par Two-dimentional subalgebra has 5 invariants.
The invariant decision is exist if all required functions are
defined from expressions for invariants. These invariants are
nominated by new functions from others invariants for this
purpose. Others invariants necessarily will be the functions of
independent variables [2].  All unknown function are defined from
the received equality. Thus, the representation of the invariant
decision turns out, which is substituted to EGD. The system of the
equations turn out which connect only invariants and new invariant
functions as a result of the substitution under the theorem of the
representation invariant variety [6]. The equations for invariants
refers an invariant's submodel.

We shall write down an invariant submodel for the considered
example.

We shall make equality from invariants (13):
${\theta}-a{\ln}{\mid}t{\mid}={\theta}_1,$
\newline $Ut-x=U_1(r,{\theta}_1), Vt=V_1(r,{\theta}_1),
Wt=W_1(r,{\theta}_1), {\rho}t^{-1}={\rho}_1(r,{\theta}_1),
pt-xb^{-1}=p_1(r,{\theta}_1).$
\newline The representation of the invariant decision is defined from these
equality:
$U=(U_1+x)t^{-1};{\:}V=V_1t^{-1};{\:}W=W_1t^{-1};$
${\rho}={\rho}_1t;{\:}p=p_1t^{-1}+x(bt)^{-1}$

The representation of the invariants decision for $S$ can be
received from the state equation: $p={\pm}{\rho}^{-1}+S
{\Rightarrow}$ $S=t^{-1}(x(b^{-1})+S_1),$ where
$S_1=p_1\pm{\rho}^{-1}$ replaced the state equation in the
invariant submodel.

The statement in EGD results to the invariant submodel$\colon$
$$D_1=(W_1r^{-1}-a){\partial}_{{\theta}_1}+V_1{\partial}_r,$$
$$D_1U_1=-({\rho}_1b)^{-1},$$
$$D_1V_1+p_{1r}{{\rho}_1}^{-1}={W_1}^2r^{-1}+V_1,$$
$$D_1W_1+p_{1{\theta_1}}({\rho}_1r)^{-1}=W_1-V_1W_1r^{-1},\eqno
(14)$$
$$D_1{\rho}_1+{\rho}_1(V_{1r}+r^{-1}W_{1{\theta}_1})=
-{\rho}_1(2+V_1r^{-1})$$
$$D_1S_1=-U_1b^{-1}.$$

Any invariant submodel can be resulted to the one of the two
initial types by the choice invariants  [7]:
\newline --   Evolutionary (time-$t$ is the invariant of the subalgebra)
$$D={\partial}_t+u_2{\partial}_s,$$
$$Du_2+b{{\rho}_1}^{-1}p_{1s}=a_1,$$
$$Dv_2=a_2,$$
$$Dw_2=a_3,\eqno (15)$$
$$D{{\rho}_1}+{\rho}_1u_{2s}=a_4$$
$$DS_1=a_5,$$
$$b>0;$$
\newline -- Stationary
$$D=u_2{\partial}_{x_1}+v_2{\partial}{y_1},$$
$$Du_2+b_1{{\rho}_1}^{-1}p_{1x_1}=a_1,$$
$$Dv_2+b_2{{\rho}_1}^{-1}p_{1y_1}=a_2,$$
$$Dw_2=a_3,\eqno (16)$$
$$D{\rho}_1+{\rho}_1(u_{2x_1}+v_{2y_1})=a_4,$$
$$DS_1=a_5,$$
$$b_1>0,{\:}b_2>0;$$
here $a_i,b,b_i$ is the coefficients of the initial types.
\newline From examining subalgebras (3) is received two submodels
 of the evolutionary type, and from others subalgebras is received
 ten submodels of the stationary type.

The canonical types of invariant submodels are tabulated (see
appendix), where:

 - 1-st column is the number of subalgebra,

 - 2-nd column is the basic system of the coordinates in which
considered the EGD,

 - 3-rd column is the initial type: S is the stationary type , E
is the evolutionary type,

 in 4-th column are given invariants,

 in 5-th column the factors of the initial type are written down.

The (14) implements to stationary initial type by replacement
\newline $r=x_1,{\:}{\theta}-a{\ln}{\mid}t{\mid},{\:}u_2=V_1,{\:}
v_2=(x_1)^{-1}W_1+a,{\:}w_2=U_1$ with factors:
$a_1=u_2+x_1(v_2+a),{\:}a_2=(v_2-a)(1-2u_2(x_1)^{-1}),{\:}
a_3=1-({\rho}b)^{-1},a_4=-{\rho}_1(u_2x_1^{-1}+2),
{\:}a_5=(1+w_2)b^{-1}+S_1,{\:}b_1=1,{\:}b_2=x_1^{-2}.$

{ The example of reduction subalgebra $ 2.9'$ to the canonical
type.}

The operators of the subalgebra are those:
\par $Y_1=aX_7-2X_{11}+X_{12},$
\par $Y_2=bx_4+X_{10}+X_{13},$ $b(\overline{\gamma})=0,{\:}
\overline{\gamma}{\neq}-1.$
\newline Invariants from independent variables look like:
$x_1=(x-b2^{-1}t^2)r^{-1},$
\newline $y_1={\theta}+a2^{-1}{\ln}{\mid}r{\mid}.$
The representation of the invariant decision enters the name
through the new invariant function: $V=V_1r^{\frac
{1}{2}},{\:}W=W_1r^{\frac {1}{2}},{\:} {\rho}={\rho}_1r^{-\frac
{1}{2}}, p=p_1r^{\frac {1}{2}}+t,{\:} U=U_1r^{\frac {1}{2}}+bt,$
where $V_1,W_1,{\rho},p_1,U_1$ depends on $x_1,y_1.$

The representation for entropy is defined from the state equation
$S=S_1r^{\frac {1}{2}}+t,$ where $S_1=p_1\pm{{\rho}_1}^{-1}.$

The substitution to the EGD results the following invariant
submodel:
\par $D_1=(U_1-x_1V_1){\partial}_{x_1}+(W_1+a2^{-1}V_1){\partial}_{y_1},$
\par $D_1U_1+{{\rho}_1}^{-1}=-b-2^{-1}V_1U_1,$
\par $D_1V_1+{{\rho}_1}^{-1}(p_{1y_1}a2^{-1}-p_{1x_1}x_1)=
W_1^2-p_1(2{{\rho}_1})^{-1}-2^{-1}V_1^2,$
\par $D_1W_1+{{\rho}_1}^{-1}p_{1y_1}=W_1V_1,$
\par $D_1{\rho}_1+{\rho}_1(U_{1x_1}-V_{1x_1}x_1+V_{1y_1}a(2r)^{-1}+
W_{1y_1})=-3V_1{{\rho}_1}2^{-1},$
\par $D_1S_1=-1-V_1S_12^{-1}.$
\newline The new invariant speeds are entered on expression for $D_1$:
\newline $U_1-x_1V_1=u_2,{\:}a2^{-1}V_1+W_1=v_2,$
$W_1-2a^{-1}V_1-x_1U_12a^{-1}$, with which we receive replacement:
$x_2={x_1}^2-a{y_1},{\:}y_2=y_1+2^{-1}a\ln{\mid}x_1{\mid},$
$u_3=2x_1u_2-av_2,{\:}v_3=(2x_1)^{-1}au_2+v_2.$ From which follows
the system (16), where:
\par
$a_1=-2x_2b-V_1(x_2U_1+2x_2(U_1-x_1V_1)-aW_1+V_1x_2(\rho)^{-1})+(2x_2-a)
(W_1^2-p_1(2{\rho}_1)^{-1}-2^{-1}V_1^2)+2(U_1-x_1V_1)^{2},$
\par $a_2=(a(2x_2)^{-1}+1)({W_1}^{2}-p_1(2{\rho}_1)^{-1}-2^{-1}{V_1}^2)-
ab(2x_2)^{-1}-aV_1U_1(4x_2)^{-1}+
(2x_2)^{-1}a(U_1-x_1V_1)V_1-2^{-1}a{x_2}^{-2}+W_1V_1,$
\par $a_3=W_1V_1-2a^{-1}({W_1}^{2}-p_1(2{\rho})^{-1}-2^{-1}{V_1}^2)+
2x_2a^{-1}(-b-2^{-1}V_1U_1)+2a^{-1}(U_1-x_1V_1)U_1,$
\par $a_4=-\frac {5}{2}{\rho}_1V_1,$
\par $a_5=-1-2^{-1}V_1S_1,$
\par $b_1=(2{x_2}^{2}2^{-1}a^2)^2+1+4{x_2}^2,$
$b_2=\frac {a^2}{4{x_2}^2}+1,$ $p_1=\pm{{\rho}_1}^{-1}+S_1.$

\centerline {5. Invariant submodel of the third rank}

The expressions for invariants are define the speed and the
pressure, but it is impossible to define density (see table,
appendix) for subalgebra $2.1''$ from optimum system (3) at $a=0$.
It is possible to build a regular partially invariant submodel in
this case.

Let is give the definition to the regular partially invariant
decisions generally.

Let us for algebra $H$ are present $I_1,..,I_k$- invariants from
independent variables and are present $J_1,..,J_l$- invariants
from dependent variables. If from invariants $J_1,..,J_l$ are
defined all dependent variables, it is possible to build an
invariant submodel of the rank $k$, nominating invariants $J_j$ by
functions from  $(I_1,..,I_k)$, i.e.
$$J_j=J_j(I_1,..,I_k), j=1,..,l. \eqno (17)$$

If it is impossible all dependent variables of invariants $J_j,$
to define, then (17) gives the representation of the regular
partially invariant decision of the rank $k,$ and defect
$\sigma,$, which is equal to number not determined independent
variables, i.e. ${\sigma}=m-l,$ where $m$ is the number of
dependent variables.

For subalgebra $2.1''$ the rank equal 3, the defect is equal 1.

Let us consider the subalgebra $2.1''$ more in detail.
\par The operators of basis are those:
\par $Y_1=\partial_t+\partial_p,$
\par $Y_2=t{\partial}_t-u{\partial}_u-v{\partial}_v-w{\partial}_w+
3{\rho}{\partial}_{\rho}+p{\partial}_p.$
\par Invariants from independent variables are : $x,y,z.$
From others invariant, specified in the table, we receive the
representation of the regular partially invariant decision..
$${ u}={\rho}^{\frac {1}{3}}{ {u_1}}(x,y,z),{\:}p=
t+{\rho}^{\frac {1}{3}}p_1(x,y,z),{\:}{\rho}={\rho}(t,x,y,z).\eqno
(18)$$ The substitution to the EGD, gives:
$$
-{\frac {1}{3}}{ {u_1}}({\rho}_t+{\rho}^{-\frac {1}{3}} {
{u_1}}{\cdot}{\nabla}{\rho})+{\rho ^{\frac {2}{3}}}[({ {u_1}}
{\cdot}{\nabla}){ {u_1}}+{\nabla}p_1]+{\frac {1}{3}}
{\rho}^{-\frac {1}{3}}p_1{\cdot}{\nabla}{\rho}=0,\eqno (19)$$
$$
{\rho}_t+{\frac {2}{3}}{\rho}^{-\frac {1}{3}}{
{u_1}}{\cdot}{\nabla}{\rho}+ {\rho}^{\frac {2}{3}} div{
{u_1}}=0.\eqno (20)$$
\newline We shall received the representation of the decision for entrophy
from the state equation
$S=t+{\rho}^{\frac {1}{3}}S_1,$ where $S_1=p_1-1.$
\par The substitution to the equation $DS=0$, gives:
$${\frac {1}{3}}S_1{\rho}^{-\frac {2}{3}}({\rho}_t+ {\rho}^{-\frac
{1}{3}}{ {u_1}}{\cdot}{\nabla}{\rho})+ {
{u_1}}{\cdot}{\nabla}S_1+1=0.\eqno (21)$$ From (20) and (21) are
follows:
$$
\frac { {u_1}{\cdot}{\nabla}\rho}{\rho}= -9{\frac {(1+{
{u_1}}{\cdot}{\nabla}S_1)}{S_1}}+ 3 div{ {u_1}}.\eqno
(22)$$ Then it is possible to find ${\rho}_t$ from the (21):
$${\frac {1}{\rho}}{\rho}_t=3{\rho}^{-\frac {1}{3}} [- div{
{u_1}}+2S_1^{-1}({ {u_1}}{\cdot}{\nabla}S_1+1)]
{\equiv}{\rho}^{-\frac {1}{3}}B({ x}).\eqno (23)$$ Replacing
$p_1$ on $S_1+1$ and substituting (23), (22) to the (19) we
receive:
$$
{\frac {1}{\rho}}{\cdot}{\nabla}{\rho}=[-\frac {1}{S_1}{
{u_1}}(1+ { {u_1}}{\cdot}{\nabla}S_1)-{\nabla}S_1-({
{u_1}}{\cdot}{\nabla}) { {u_1}}]{\frac
{3}{S_1+1}}{\equiv}{ A}({ x}).\eqno (24)$$ By the
substitution (24) to (22), we exclude $\rho$:
$$
(\frac {{ u_1}^2}{S_1+1}-3)({ {u_1}}{\cdot}{\nabla}S_1+1)+
\frac {S_1}{S_1+1}({ {u_1}}{\cdot}{\nabla}) (S_1+\frac
{1}{2}{ {u_1}}^2)=0.
$$
Equating the mixed derivative functions $\ln{\rho}$ from (23),
(24), we receive ${\nabla}B={\frac {1}{3}}B{ A},$ $
rot{ A}=0.$ From the last equality follows, that ${
A}=\nabla{\varphi}$ and ${\nabla}(3{\ln}B-{\varphi})=0
{\Rightarrow} 3{ ln}B-{\varphi}=0 {\Rightarrow} B=e^{{\frac
{1}{3}}{\varphi}}.$ From (24) follows ${\rho}=b(t)e^{\varphi}.$
Then from (21) we receive $b'=b^{\frac {2}{3}}.$

The integration gives $b=(\frac {t}{3})^3,$ where constant of the
integration is made by zero with the help of carries on $t$ and on
$p,$ admitted by EGD.

So, density as $\rho=t^3{\rho}_1(x,y,z)$, i.e. is the
representation of the invariant decision for one-dimensional
subalgebra $Y_2.$

Thus, the is a reduction of the partially invariant decision to
the invariants:
$$
({ u_1}\cdot{\nabla}){ u_1}+{\rho_1}^{-1}\cdot{p_1}={
u_1},
$$
$$
{ u_1}\cdot{\nabla}\rho_1+\rho_1 div{
u_1}=-3\rho_1,\eqno (25)$$
$$
{ u_1}\cdot{S_1}=-S_1,
$$
where $S_1=p_1-{\rho_1}^{\frac {1}{3}}$, $S=tS_1.$

\centerline { Literature }

[1] Ovsyannikov L.V. Group properties of differential equations. -
Novosibirsk: SO AN SSSR, 1962. - 240 p.

[2] Ovsyannikov L.V. Group analysis of differential equations. -
M.: Nauka, 1978. - 400 p.

[3] Khabirov S.V. The invariant decisions of the rank 1 in gas
dynamics // Works of the international conference "Modeling,
calculation, designing in conditions of uncertainty". - Ufa:
USATU, 2000, P. 104-115.

[4] Khabirov S.V. The optimum systems of the subalgebras, admitted
by the equations of gas dynamics. - Ufa: Institute of the
mechanics USC of RAS, 1998.-33р.

[5] Stanukovich K.P. The unsteady movements of continous
environment. - M.: GITTL, 1955. - 804 p.

[6] Gunter N.M. The integration first order partial differential
equations. -  L.M.: GTTI, 1934. - 359 p.

[7] Khabirov S.V. The reduction of invariant submodel gas dynamics
to canonical form // Mat. Zametki. - 1999. - V. 66. - N. 3. - pp.
439 - 444.

\centerline {Appendix}

2.1'. \  C.S.: D,\ Type: S;

Invariants: \ $ y,z,tv,tw,tu-x+\ln\mid t \mid, \rho t^{-1},
tS-xa^{-1}+a^{-1}\ln\mid t\mid$;

 Submodel (16): \ $a_1=u_2, a_2=v_2,
         a_3=1-(a\rho_1)^{-1}, a_4=-2\rho_1,
         a_5=S_1+a^{-1}(1-w_2), b_1=b_2=1$;

2.2'. \  C.S.: D,\ Type: S;

Invariants: \ $ y,z,tv,tw,tu-x,\rho t^{-1}, tS-xa^{-1}$;

Submodel (16): \ $a_1=u_2, a_2=v_2,
         a_3=-(a\rho_1)^{-1}, a_4=-2\rho_1,
         a_5=S_1+a^{-1}w_2, b_1=b_2=1$;

2.3'. \  C.S.: D,\ Type: S;

Invariants: \ $x-ayb^{-1}-\ln\mid t \mid, z, tu-ab^{-1}tv-1, tw,
ab^{-1}(tu-ayb^{-1}-1)+tv-y, {\rho}t^{-1}, tS-ya^{-1}$;

Submodel (16): \ $a_1=u_2+a(b^2\rho_1)^{-1}+1, a_2=v_2\rho_1,
a_3=ab^{-1}(u_2+1)-(b\rho_1)^{-1}, a_4=-3\rho_1,
a_5=S_1-ba^{-2}w_2+a^{-1}u_2, b_1=a^2b^{-2}+1, b_2=1$;

2.4'. \  C.S.: D,\ Type: E;

Invariants: \ $t,z^{-1}(y-atx), z^{-1}(v-ax-atu-sw),
z^{-1}[(a^2t^2+s^2)(v-ax)+atu+sw],
z^{-1}((a^2t^2+s^2)w+s(v-ax)-atsu)\rho{z}, z^{-1}(S-x)$;

Submodel (15): \
$a_1=-2(1+ a^2{t}^2+S^2)^{-1}[u_2(a^2t^2-stu_2)+
w_2(tu_2-s)]=at(\rho_1)^{-1}-sp_1(\rho_1)^{-1},
a_2=(1+a^2t^2+s^2)^{-1}[2av_2-u_2a^3t^2-2asw_2+
(v_2+u_2)(2a^2t+su_2)-w_2v_2+ sv_2u_2+w_2u_2+s(u_2)^2]+
at(\rho_1)^{-1}-sp_1(\rho_1)^{-1}, a_3=(\rho_1)^{-1}
(p_1(1+a^2t^2+ ats)+(1+a^2t^2+s^2)^{-1}
[2a^2t^2w_2-(w_2\!-su_2)^2t-2sv_2+ as^2w_2]+(u_2)^2,
a_4=-2\rho((w_2-su_2)(1+a^2t^2+s^2)^{-1},
a_5=[w_2(S_1-s)+v_2-u_2(a^2t+S_1s)](1+a^2t^2+s^2)^{-1},b=(1+a^2t^2+
s^2), p_1={\rho_1}^{-1}+S_1$;

2.5'. \  C.S.: D,\ Type: E;

Invariants: \ $t, \theta +a\ln\mid t\mid, r^{-1}(aV+W),r^{-1}U,
ar^{-1}(Wa-V), pr, r^{-1}(S-x)$;

Submodel (15): \ $a_1=-
a(\rho_1)^{-1}(S_1+(\rho_1)^{-1})+
(a(a^2+1))^{-1}({w_2}^2-a^2{u_2}^2+2u_2w_2, a_2=-(\rho_1)^{-1}-
v_2(a(a^2+1))^{-1}(au_2-w_2), a_4=-\rho_1(a(a^2+
1))^{-1}(a_2u_2-w_2), a_5=-v_2, b=a^2+1$;

\end